\documentclass[aps,prl,twocolumn,showpacs,superscriptaddress]{revtex4-1} 

\usepackage[dvips]{graphicx}
\usepackage{color}
\usepackage{psfrag}
\usepackage{ulem}

\usepackage{amsthm}
\usepackage{amsmath}
\usepackage{amsfonts}
\usepackage{amssymb}

\usepackage[utf8]{inputenc}
\usepackage{mathptmx}
\usepackage{t1enc}

\def\ad{\mathrm{ad}}

\def\tJ{\tilde{J}}
\def\mO{\mathcal{O}}
\def\mR{\mathcal{R}}
\def\mL{\mathcal{L}}
\def\ct{c_{\theta}}
\def\st{s_{\theta}}
\newcommand{\Hm}[1]{\tilde{H}_{#1}}

\newcommand{\tr}{{\text{Tr}}}

\newcommand{\expv}[1]{\left\langle {#1} \right\rangle}

\begin{document}
\title{Replica resummation of the Baker-Campbell-Hausdorff series}
\author{Szabolcs Vajna}
\affiliation{Department of Physics, Faculty of Mathematics and Physics, University of Ljubljana,  SI-1000 Ljubljana, Slovenia}
\affiliation{Department of Physics and BME-MTA Exotic  Quantum  Phases Research Group, Budapest University of Technology and Economics, 1521 Budapest, Hungary}
\author{Katja Klobas}
\affiliation{Department of Physics, Faculty of Mathematics and Physics, University of Ljubljana,  SI-1000 Ljubljana, Slovenia}
\author{Toma\v{z} Prosen}
\affiliation{Department of Physics, Faculty of Mathematics and Physics, University of Ljubljana,  SI-1000 Ljubljana, Slovenia}
\author{Anatoli Polkovnikov}
\affiliation{Department of Physics, Boston University, 590 Commonwealth Ave., Boston, MA 02215, USA}

\begin{abstract}
We developed a novel perturbative expansion based on the replica trick for the Floquet Hamiltonian governing the dynamics of periodically kicked systems  where the kick strength is the small parameter. The expansion is formally equivalent to an infinite resummation of the Baker-Campbell-Hausdorff series in the un-driven (non-perturbed) Hamiltonian, 
while considering terms up to a finite order in the kick strength. As an application of the replica expansion, we 
analyze an Ising spin 1/2 chain periodically kicked with magnetic field of strength $h$, which has both longitudinal and transverse components.
We demonstrate that even away from the regime of high frequency driving, the heating rate is nonperturbative in the kick strength bounded from above by a stretched exponential: $e^{-{\rm const}\,h^{-1/2}}$. This guarantees existence of a very long pre-thermal regime, where the dynamics is governed by the Floquet Hamiltonian obtained from the replica expansion.
\end{abstract}
\maketitle

\normalem 

\emph{Introduction.---}
Time-periodic modulation of interactions is a powerful tool to engineer properties of materials in both artificial and condensed matter systems \cite{BukovAdvPhys2015}. In particular, high frequency driving is the cornerstone of various experiments and proposals inducing interactions such as spin-orbit coupling \cite{AndersonPRL2013}, artificial gauge fields for uncharged particles \cite{MiyakePRL2013,AidelsburgerPRL2013}; it has been applied to dynamically tune or suppress hopping amplitude in optical lattices \cite{ZenesiniPRL2009}, and also to change topological properties of materials \cite{JotzuNAT2014,rechtsman,GoldmanNatPhys2016}.

Given a periodic driving protocol, however, determining the Floquet Hamiltonian that governs the stroboscopic evolution is usually a highly non-trivial task. Except for some special integrable cases \cite{WilcoxJMP1967,VanBruntJPA2015,GritsevArxiv2017}, one is compelled to apply approximate methods, e.g. variants of high frequency expansion (Magnus \cite{Magnus1954}, van Vleck \cite{RahavPRA2003} or Brillouin-Wigner \cite{MikamiPRB2016} expansions). These provide a local effective Hamiltonian in each order of the expansion, however, until recently, not much had been known about the convergence properties of these series. A conjecture based on the generalization of the eigenstate thermalization hypothesis suggests that generic closed periodically driven systems heat up in the thermodynamic limit, i.e. they approach a completely structureless, infinite temperature steady state \cite{PonteAOP2015,LucaPRX2014,LazaridesPRE2014}. The convergence of the expansions of the effective Hamiltonian is intimately related to heating. Recently upper bounds on heating had been reported in the linear response regime \cite{AbaninPRL2015} and for the Magnus expansion \cite{MoriPRL2016,KuwaharaAoP2016}, with the central result that the heating is \emph{at least} exponentially suppressed in the driving frequency for periodically driven models characterized by local Hamiltonians with bounded energy spectrum. This theoretical finding implies that one can engineer nontrivial phases of matter, which remain stable for the experimentally relevant timescales. 
In some situations heating seems to be either absent completely or remain well below exponential bounds~\cite{prosen_98a, ProsenPRE02,LucaAoP2013, citro_15}. Another recent theoretical work showed that nontrivial non-equilibrium Floquet phases can be stabilized by weak coupling to environment~\cite{lenarcic_17}.

\begin{figure}[t!]
\centering
\includegraphics[width=8.5cm]{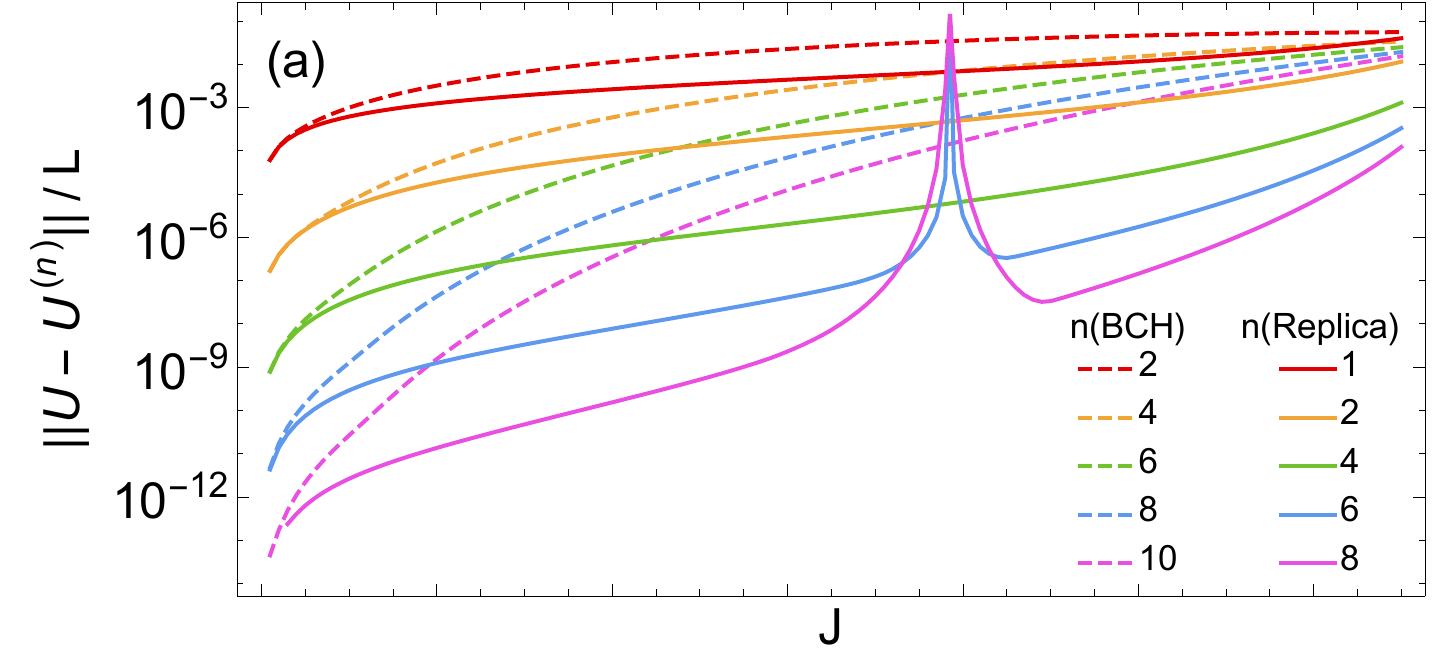}\\
\includegraphics[width=8.5cm]{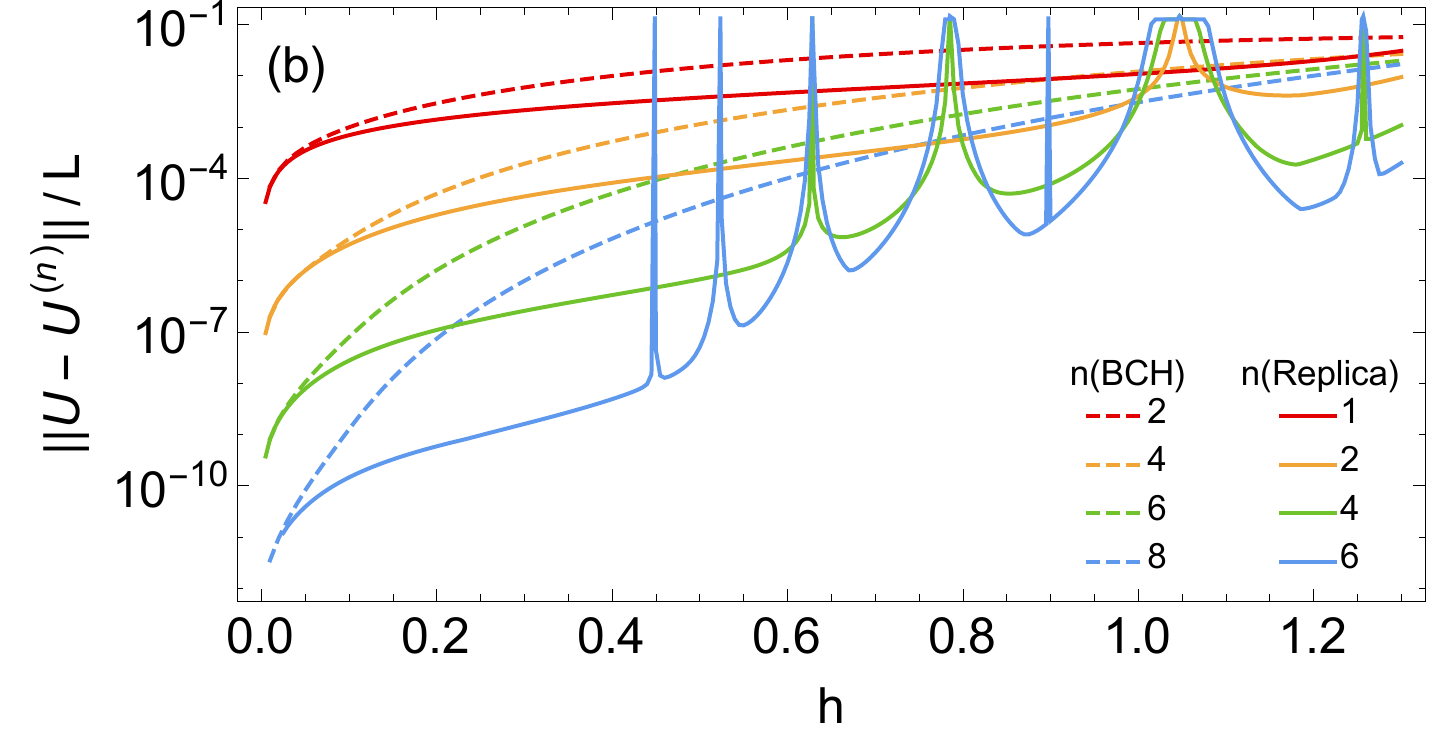}
\caption{The replica expansion (solid lines) beats the traditional Magnus (BCH) expansion (dashed lines) by several orders of magnitude away from the narrow resonances at rational fractions of $\pi$. The performance of the expansions is measured by the $l_2$ distance of the exact ($U$) and the approximate time evolution operator ($U^{(n)}=\exp(-i H_F^{(n)})$) within a single time-period, in the kicked tilted field Ising model defined in Eq.~\eqref{eq:Ham_TFIM}. In the top (bottom) panel the magnetic field (Ising interaction) is considered as the periodic kick. The orders are chosen such that for the matching colors the replica expansion contains all the nested commutators appearing in the corresponding order of the BCH expansion. The curves were obtained by exact diagonalization using the QuSpin package \cite{WeinbergQuSpin} on a system of $L=16$ sites and kick strength (a) $h=0.1$ (b) $J=0.1$. The direction of the magnetic field is defined by $\ct=0.8$, $\st=0.6$.}
\label{fig:replica_vs_BCH}
\end{figure}

One of the most studied driving protocols is a time-periodic sequence of sudden quenches between different Hamiltonians \cite{GoldmanPRX2014,LucaAoP2013},
which is interpreted as kicked dynamics if one of the time intervals on which the Hamiltonians act is much shorter than the other, or equivalently, the strength of one of the Hamiltonians is smaller than the other. Such protocols naturally appear e.g. in the context of digital quantum simulation in trapped ions~\cite{CiracPRL1995,lanyon_11,BlattNatPhys2012} and are frequently realized in other experimental platforms~(see e.g. Refs.~\cite{neil_16, hoang_13}).

In the present work a novel expansion for the effective Floquet Hamiltonian is introduced for periodically kicked systems, which clearly outperforms the traditional high frequency expansions in a wide parameter range, as illustrated using two examples shown in Figure~\ref{fig:replica_vs_BCH}. Our approach uses the replica trick to calculate the logarithm of the time evolution operator describing a single period. The small parameter is the kick strength and we do not assume high frequency driving. The Magnus expansion is equivalent to the Baker-Campbell-Hausdorff (BCH) series in periodically kicked systems, and the replica expansion can be thought of as an infinite resummation of the BCH formula.  As such, the possible applications of the replica expansion can reach far beyond periodically driven systems, including the theory of differential equations \cite{Magnus1954}, Lie group theory \cite{vv1997foundations}, analysis of NMR experiments \cite{UntidtPRE2002} or estimation of Trotterization errors in various numerical integration schemes. As an application of the replica expansion, we establish a conjecture for a non-perturbative -- stretched exponential $\exp(-{\rm const}\,h^{-1/2})$ in the kick strength $h$ -- upper bound for the heating rate in the kicked tilted field Ising model. Our conjecture supplements the bounds introduced in Refs.~\cite{AbaninPRL2015,MoriPRL2016,KuwaharaAoP2016}, which address only the regime of high frequency driving. 

\emph{Kicked dynamics.---}
The effective Floquet Hamiltonian evolving the kicked system is given by the logarithm of the stroboscopic time evolution operator over one period $U=U_0 U_1=e^{-i J H_0}e^{-i h H_1}$ as 
\begin{align} \label{eq:H_F_def}
H_F=i\log(e^{-i J H_0}e^{-i h H_1}) \,,
\end{align}
where we incorporated the time intervals of the Hamiltonians $H_{0,1}$ to the coupling constants $J,h$. The BCH formula provides a series expansion for $H_F$ assuming both $J$ and $h$ are small as 
\begin{align}
H_F=&J H_0+h H_1-i J h \frac{1}{2}[H_0,H_1]-\nonumber\\ &Jh\frac{1}{12}(J[H_0,[H_0,H_1]]+h[H_1,[H_1,H_0]]) +\dots
\end{align}
In the usual setup of kicked systems, where $J\gg h$, that is, at intermediate frequencies and weak kick strengths, the terms with low order in $H_1$ but high order in $H_0$ are not negligible. A series expansion in the small parameter of $h$ could be obtained formally by a resummation of the BCH series in $H_0$. The first order resummation is well known \cite{ScharfJPA1988,LucaAoP2013},
\begin{align} \label{eq:BCH_resum_ord1}
H_F=J H_0+\frac{-i J \ad_{H_0}e^{-i J\ad_{H_0}}}{e^{-i J\ad_{H_0}}-1}h H_1+\mO(h^2)
\end{align}
where $\ad_X(Y)=[X,Y]$ is the Lie derivative. However, to the best knowledge of the authors, closed form expressions for the infinite resummation in higher orders of $H_1$ have not yet been reported in the literature.

\emph{Replica expansion.---}
We tackle this problem by constructing a series expansion in $h$ in Eq.~\eqref{eq:H_F_def}. Because of the noncommutativity of $H_0$ and $H_1$, the higher order derivatives of the logarithm of the time evolution operator cannot be obtained easily.  To circumvent this obstacle, we apply the replica trick to express the logarithm,   
\begin{align}\log U=\lim_{\rho\rightarrow 0}\frac{1}{\rho}(U^{\rho}-1).\end{align} 
This idea has been proven to be uniquely useful in various fields of science, such as in the statistical physics of spin glasses \cite{Mezard1987spin}, machine learning \cite{engel2001statistical}, and also in calculation of the entanglement entropy \cite{CardyJSP2008}. 
Assuming that the replica limit $\mL(\bullet)\equiv\lim_{\rho\rightarrow 0}\frac{1}{\rho}(\bullet)$ commutes with the differentiation, the series expansion of the Floquet Hamiltonian in Eq.~\eqref{eq:H_F_def} reads 
\begin{align}
H_F^{(n)}=\sum_{r=0}^{n} h^r \Gamma_r\,,
\end{align}
with 
$
\Gamma_0=J H_0 $ and 
$\Gamma_r=\mL\, \frac{1}{r!}\partial^r_h U^\rho
$.
The derivatives of the powers of the time evolution are easy to calculate at integer values of the replica index $\rho$, and the replica limit is taken following an analytical continuation to arbitrary real values. After evaluating the derivatives and neglecting the prefactor $U_0^{\rho}$, which approaches the identity operator in the replica limit, the $r^{\text{th}}$ order correction reads 
\begin{align}\label{eq:Gamma_noncomm}
 \Gamma_r=\frac{(-i)^{r-1}}{r!}\hspace{0.5em}  \mL\hspace{-1em} \sum_{0\leq m_1\leq \dots m_r<\rho} \hspace{-1em} \Hm{m_r}\Hm{m_{r-1}}\dots \Hm{m_1} c_{m_1 \dots m_r}
\end{align}
where $\Hm{m}=U_0^{-m} H_1 U_0^m$ and $c_{m_1 \dots m_r}=\frac{r!}{n_0!n_1!\dots}$ is the multinomial coefficient with $n_s$ being the number of indices taking value $s$.
Our method provides a remarkably simple derivation of the known first order term in Eq.~\eqref{eq:BCH_resum_ord1} giving a certain degree of confidence in the replica expansion (see \cite{suppmat}).

It is not clear directly from Eq.~\eqref{eq:Gamma_noncomm} if higher order terms can be expressed as nested commutators (multiple Lie derivatives), which would be expected from the resummation of the BCH series. One can show that the corrections to the Floquet Hamiltonian can be represented as sums of terms containing the commutators and extra terms, which vanish in the zero replica limit. For example at second order 
\begin{align}
\Gamma_2=\frac{-i}{2} \mL \Big\{\sum_{0 \leq m_1 \leq  m_2<\rho} [\Hm{m_2},\Hm{m_1}]+ \big(\sum_{0 \leq m_1 <\rho} \Hm{m_1}\big)^2\Big\} \,,
\end{align}
where the second term is proportional to $\rho^2$, assuming that the replica trick works at first order, that is, $\sum \Hm{m}\sim \rho$. Similar transformations that produce the nested commutator expression are given in the supplemental material up to the $5^{\text{th}}$ order.

The resulting expansion of the replica Floquet Hamiltonian expressed in terms of nested commutators reads
\begin{subequations}
\begin{align}
 \Gamma_0&=J H_0\\
 \Gamma_1&=\mL \sum_{0\leq m < \rho} \Hm{m}\\
 \Gamma_r&=\frac{(-i)^{r-1}}{r!}\hspace{0.2em}  \mL\hspace{-1em} \sum_{0\leq m_1\leq \dots m_r<\rho} \hspace{-1em} [\Hm{m_r},\dots [\Hm{m_2},\Hm{m_1}]]] c_{m_2 \dots m_r}
\end{align}
\label{eq:Gamma_comm}%
\end{subequations}
where $c_{m_2 \dots m_r}=\frac{(r-1)!}{n_0!n_1!\dots}$. The expansion can be constructed similarly for different initial phases of the driving, $U'=e^{-i J H_0 (1-\varphi)}e^{-i h H_1} e^{-i J H_0 \varphi}=e^{-i H_F'}$, which results in the same equations as Eq.~\eqref{eq:Gamma_comm} except for a simple substitution $\Hm{m_i}\rightarrow \Hm{m_i+\varphi}$.
Having established the first main result of this Letter, we now demonstrate its performance in the example of the kicked Ising model in a tilted field \cite{ProsenPRE02}.

\emph{Kicked Ising model.---} 
The time evolution is characterized by time-periodic quenches between the Hamiltonians $H_{0,1}$,
\begin{subequations}
\begin{align}
 H_0&=\sum_i \sigma_i^z \sigma_{i+1}^z\\
 H_1&=\sum_i \ct \sigma_i^x+ \st \sigma_i^z\,, 
\end{align}
\label{eq:Ham_TFIM}%
\end{subequations}
where $\ct$ and $\st$ are shorthand notations for $\cos\theta$ and $\sin\theta$. The purpose of introducing the tilt angle is to break the integrability of the model  at  $\theta= 0$ and $\pi/2$ \cite{ProsenJPA1998}. Figure~\ref{fig:replica_vs_BCH} shows the performance of the replica expansion for the kicked Ising model in two different limits: when the kick parameter is the magnetic field $h$, or the Ising interaction $J$. 
The spectral norm of the difference between the approximate and exact time evolution operators, $\Delta_n=\|U-U^{(n)}\|$ shown in Figure~\ref{fig:replica_vs_BCH}, bounds the accuracy of the expansion for the dynamics of \emph{any observable} $A$, as $|\expv{A}(t)-\expv{A}_n(t)|\leq 2 t \|A\| \Delta_n +\mO(\Delta_n^2) $, where $\expv{A}(t)$ ($\expv{A}_n(t)$) is the expectation value of the observable following $t$ periods with respect to the exact (approximate) time evolution, starting from an arbitrary initial state \footnote{Although $\Delta_n$ scales linearly with the system size $L$, this accounts for expectation values of nonlocal operators. The accuracy of the dynamics of local observables is independent of system size \cite{suppmat}.}. 
As the two cases, kicking with $H_0$ or $H_1$, show qualitatively similar behavior, we discuss here only the replica expansion for kicking magnetic field.

The time evolution of kicking Hamiltonian $H_1$ with respect to unperturbed dynamics $H_0$ for $m$ periods reads explicitly
\begin{align}
 \Hm{m}= \sum_i & \st \sigma_i^z-\ct \Big(\frac{1}{2}\sin(4mJ)(\sigma_{i-1}^z\sigma_{i}^y+\sigma_{i}^y\sigma_{i+1}^z)+ \nonumber
 \\&  \sin^2(2mJ)\sigma_{i-1}^z\sigma_{i}^x\sigma_{i+1}^z -\cos^2(2mJ)\sigma_i^x\Big) \,,
\end{align}
and is the main building block of the replica expansion. The computation of the nested commutators of these objects is trivial, following which one has to deal with the multiple sums and the replica limit. It is convenient to separate the operator part of the expansion from the replica coefficients by expressing $\tilde{H}_m$ in Fourier harmonics as $\Hm{m}=\mO_0+e^{i 4 J m} \mO_1+e^{-i 4 J m}\mO_{-1}$, where
\begin{subequations}
\begin{align} \label{eq:H_m_Fourier}
\mO_0&=\sum_i \st \sigma_i^z+\frac{\ct}{2}(\sigma_i^x-\sigma_i^z \sigma_{i+1}^x\sigma_{i+2}^z)\\
\hspace{-0.5em}\mO_{\pm 1}&=\sum_i \frac{\ct}{4}[\sigma_i^{x}+\sigma_i^z \sigma_{i+1}^x\sigma_{i+2}^z\pm i (\sigma_i^z\sigma_{i+1}^y+\sigma_i^y\sigma_{i+1}^z)],
\end{align}%
\end{subequations}
which brings us to the simplest formulation of the replica expansion,
\begin{align} 
 \Gamma_1&=\sum_{x_1}\mR_{x_1}\mO_{x_1}\\
 \Gamma_{r}&=\frac{(-i)^{r-1}}{r!}\sum_{x_1,x_2,\dots,x_r}\mR_{x_1 x_2 \dots x_r} [\mO_{x_r},\dots [\mO_{x_2},\mO_{x_1}]] \label{eq:Gamma_r_mR}
\end{align}
where $x_i\in\{0,\pm 1\}$ and we introduced the replica sum
\begin{align}  \label{eq:replica_sum}
 \mR_{x_1 x_2 \dots x_r}&=\mL \hspace{-1.5em} \sum_{0\leq m_1 \leq \dots m_r < \rho} \hspace{-1.5em}  e^{i 4J m_1 x_1}e^{i 4J m_2 x_2}\dots e^{i 4J m_r x_r} c_{m_2 \dots m_r}\,.
\end{align}
These sums are evaluated gradually (with attention to the combinatorial factors) as 
\begin{align}\label{eq:repl_sum_part}
\sum_{m_j=0}^{m_{j+1}-1}\hspace{-0.5em} m_j^{y} e^{i \tJ m_j}&=\Big(\frac{\partial}{i \partial_{\tJ}}\Big)^{y}\frac{e^{i \tJ m_{j+1}}-1}{e^{i \tJ}-1} \,,
\end{align}
with $m_{r+1}=\rho$, and $\tilde{J}$ is an integer multiple of $4J$. The prefactor $m^y$, $0\leq y\in \mathbb{N}$ may arise from the previous sum with respect to $m_{j-1}$, e.g. from the sum of constant terms. This way of evaluating the sums already defines the analytical continuation to arbitrary real values of $\rho$, allowing one to take the replica limit $\mL$. 
This analytical continuation leads to $\mL e^{i \tJ \rho}-1=\log e^{i \tJ}=i \tJ$, which tries to enforce a Floquet Hamiltonian continuous in $J$ at a price of breaking the periodicity $H_F(J)=H_F(J+2\pi)$. Alternatively, one can choose a different branch of the logarithm, e.g. which folds $J$ into the interval $(-\pi,\pi]$ by applying a different analytical continuation \cite{suppmat}. This ambiguity in choosing the branch of the logarithm can be potentially used to further improve the expansion, restore the periodicity in $J$ and eliminate divergences which are discussed below.

The sum is especially simple in the first order correction: {$\mR_0=0$, $\mR_{\pm 1}=2J(\cot 2J \mp i)$}, yielding 
 \begin{align}\nonumber
 \Gamma_1=\sum_i & a_{+} \sigma_i^x+ a_{-}\sigma_{i-1}^z\sigma_{i}^x \sigma_{i+1}^z+\st \sigma_i^z+\\ &\ct J (\sigma_{i-1}^z\sigma_{i}^y+\sigma_{i}^y\sigma_{i+1}^z)
 \label{eq:Gamma_1_TFIM}
\end{align}
with $a_{\pm}=\ct(J\cot 2J \pm 1/2)$. 

The second order correction is written in a compact form by noticing that  $\mO_{\pm 1}^{\dagger}=\mO_{\mp 1}$ and $\mR_{x_1,x_2}^{*}=\mR_{-x_1,-x_2}$,
\begin{align}
\hspace{-0.58em} \Gamma_2=\frac{-i}{2}\{(\mR_{10}-\mR_{01})[\mO_0,\mO_1]+\mR_{1-1}[\mO_{-1},\mO_1]\}+\text{h.c.}
\end{align}
The replica coefficients are evaluated as
\begin{align}
 \mR_{10}-\mR_{01}&=(1-2J \cot 2J)(1+i \cot 2J)\\
 \mR_{1-1}&=\frac{1}{2}-i\frac{\sin 4J-4J}{4\sin^2 2J}\,,
\end{align}
which finally yields
\begin{align} \nonumber
\Gamma_2=&\sum_i \ct a_{-} (\sigma_i^y \sigma_{i+1}^x \sigma_{i+2}^z+\sigma_i^z \sigma_{i+1}^x \sigma_{i+2}^y) - \\ \nonumber 
&\st a_{-} [\sigma_i^y+\sigma_i^z \sigma_{i+1}^y \sigma_{i+2}^z+\cot{2J}(\sigma_i^x \sigma_{i+1}^z +\sigma_i^z \sigma_{i+1}^x )]+
\\&(b+c)\sigma_i^z \sigma_{i+1}^x\sigma_{i+2}^x \sigma_{i+1}^z -b \sigma_i^y \sigma_{i+1}^y-c \sigma_i^z \sigma_{i+1}^z\,.
\end{align}
The coefficients are $b=\frac{\ct^2}{8}\frac{4J\cos 4J-\sin 4J}{\sin^2 4J}$ and $c=\frac{\ct^2}{4}\frac{4J-\sin 4J}{\sin^2 4J}$. The higher order corrections can be  calculated similarly \cite{suppmat}.

Notice that the first order correction diverges near $J_{k,1}=k\pi/2$, which was identified as a signal of a heating (or nonergodicity--ergodicity) transition in a different spin model \cite{LucaAoP2013}, similar to the divergence of high-temperature expansion signaling phase transition in statistical physics. 
The source of this divergence is easily identified as the zero of the denominator in Eq.~\eqref{eq:repl_sum_part}. Similar to the high frequency expansion, the higher order corrections become less and less local due to the increasing number of commutators.  
The degree of divergence at $J_{k,1}$ also increases with the order, as the denominators from the consecutive sums become multiplied, and it can also increase because of the derivative in Eq.~\eqref{eq:repl_sum_part}, leading to a divergence $\sim |J-k \pi/2|^{-r}$ at the $r^{\text{th}}$ order of expansion. These divergences restrict the convergence radius of the expansion. 
A finite convergence radius would imply no heating in the domain of convergence and would suggest the existence of a heating transition in the parameter $J$. However, at order $r$, in addition to the $r^{\text{th}}$ order divergence at $J=k\pi/2$, additional lower order divergencies may appear at $J_{k,m}=k\pi/2m$, $m=1\dots r$. Consider e.g. the replica sum $\mR_{11}=2J (\cot 4J-i)$ appearing in the second order expansion, which diverges at $J_{k,2}=k \pi/4$. Many of these possible divergences do not enter the expansion because of the vanishing commutators in the operator part or due to cancellations, e.g. $[\mO_1,\mO_1]=0$, $[\mO_1,[\mO_1,\mO_0]]=0$, etc. For instance, the divergence at $k \pi/4$ only appears at the $5^{\text{th}}$ order of the expansion, see Figure~\ref{fig:replica_vs_BCH}. In spite of the cancellations we conjecture that new divergences keep appearing in increasing orders similar to the dual case with interaction kicks (Figure~\ref{fig:replica_vs_BCH}(b)), and the expansion blows up near every rational fraction of $\pi/2$ (similar to KAM series).

In spite of the divergences at high orders of the expansion, it can provide a very accurate estimate of the Floquet Hamiltonian at low orders. Moreover, the divergences determine the order at which to stop the expansion. That is, given a fixed $J$, similar to the method introduced in Ref.~\cite{MoriPRL2016}, one can introduce an optimal order of expansion $n^*$, up to which the corrections increase the accuracy of the approximation of the Floquet Hamiltonian. 

It is natural to assume that the width of the resonances is proportional to the small parameter $h$, which is further supported by the analysis of the magnitude of the corrections $\Gamma_r$
\footnote{We give a more rigorous discussion and estimate about the magnitudes of $\Gamma_r$ in a subsequent publication.}. As an illustration, we give the scaling of the Hilbert-Schmidt norm $\|\Gamma_r\|_{\rm HS} = \sqrt{\tr\, \Gamma^\dagger_r \Gamma_r}
\sim f^r_J(J-\pi/4)$ near the resonance $\pi/4$,
\begin{align} \label{eq:div_coeff}
f^r_{\pi/4}(\delta J)&=\frac{c_{\pi/4}(r)}{\delta J^{r-4}}+\mathcal{O}(\delta J^{-(r-5)})\,.
\end{align}
The $r$ dependence of the prefactor is illustrated in \cite{suppmat}. Up to the highest order we had access to, we found $c_J(r)$ to decrease with $r$. For our purposes it is enough to assume that it grows at most exponentially $\lesssim \alpha^r$, and we expect that at high orders this exponential growth indeed appears as the asymptote of $c_J(r)$. Then the series $\sum_r \|\Gamma_r\|h^r$ diverges for $\delta J< \alpha h$, which gives the width of the resonances.
The optimal order of the expansion is hence estimated by the maximal order at which the closest resonance is located further than $\sim \alpha h$.
As the resonances appear at the rational fractions of $\pi/2$, $J=\frac{k\pi}{2m}$, where $m=1,\dots, n$ at the $n^{\text{\text{th}}}$ order of the expansion, the question is how far one can get in the expansion without having a resonance approaching a fixed $J$.

Rational approximation of irrational numbers has been thoroughly studied in the mathematical literature \cite{khinchin1997continued}, and is the cornerstone of the KAM theorem in classical dynamical systems, where the stability of the 
(quasi)periodic motion to integrability-breaking perturbations depends on the irrationality of the corresponding frequencies.
The irrationality of a number is defined by how difficult it is to approximate by rational numbers. The number $x$ is of type $(K,\nu)$ if it satisfies
$|x-p/q|>K q^{-\nu}$
for all integer pairs $(p,q)$ \cite{arnold2012geometrical}.  For example, the most irrational number in this sense is the golden ratio, which is of type $(1/\sqrt{5},2)$. Such badly approximable numbers are generic in the sense that for any $\nu>2$, almost all irrational numbers $x$ are of type $(K,\nu)$ for some $K$ \cite{khinchin1997continued,arnold2012geometrical}. In the following we choose a $J$ for which $\frac{2J}{\pi}$ is of type $(K,\nu)$, such that
\begin{align}
 \left|J-\frac{k\pi}{2 m}\right|=\frac{\pi}{2}\left|\frac{2J}{\pi}-\frac{k}{m}\right|>\frac{\pi}{2}\frac{K}{m^{\nu}}\,.
\end{align}
Hence $J$ is not affected by any resonances as long as $n<n^*$
\begin{align}
 n^*=\left(\frac{\pi K}{2\alpha h}\right)^{\frac{1}{\nu}} \,,
\end{align}
which we set as the optimal order of expansion. By the construction of the expansion,
\begin{align}
 \|U-e^{-i H_{F}^{(n)}}\|\sim h^{n+1} \,,
\end{align}
which gives
\begin{align}
 \|U-e^{-i H_{F}^{(n^*)}}\|\sim h^{n^*+1}\sim h^{\frac{C}{h^{1/\nu}}}\lesssim e^{-\frac{C'}{h^{1/2-\epsilon}}}
\end{align}
at the optimal order with some constants $C$, $C'$ and arbitrary $\epsilon>0$, by choosing $\nu$ close enough to $2$. Consequently, the Floquet Hamiltonian in the optimal order is conserved for stretched exponentially long times in the inverse kick strength, and, if the steady state is the infinite temperature ensemble, it is approached at least stretched exponentially slowly.
In the above analysis we gave an estimate for the accuracy of the replica expansion. We leave a more rigorous mathematical analysis, similar to the ones in Refs.~\cite{MoriPRL2016,KuwaharaAoP2016}, to future work.

\emph{Conclusion.---}
We have developed a novel expansion applicable to periodically driven systems where the driving consists of sudden quenches between different Hamiltonians. The expansion takes into account all orders in one of the Hamiltonians and is perturbative in the other. As such, it is an infinite resummation of the BCH formula, whose coefficients can be reproduced by taking the derivatives of the terms in the replica expansion \cite{suppmat}. We demonstrated that, similar to the high frequency expansions, the replica expansion is  asymptotic for systems with unbounded Hamilton operators, that is, it may not converge, but performs very well when evaluated at an optimal order. The expansion suffers from resonances near rational frequencies, whose avoidance determines the optimal order of expansion. It is an interesting question whether these resonances have a physical meaning or   
are just an artifact of the expansion, and whether one could remove the resonances by a proper choice of analytical continuation in the replica trick.

\begin{acknowledgments}
This research has been supported by the grants P1-0044 and N1-0025 of Slovenian Research Agency (ARRS), Hungarian-Slovenian (OTKA/ARRS) bilateral  grant N1-0055, ERC AdG grant OMNES, and grants by the Hungarian National Research, Development and Innovation Office - NKFIH K119442, SNN118028. A.P. was supported by NSF DMR-1506340, ARO W911NF1410540 and AFOSR FA9550-16-1-0334.	
\end{acknowledgments}

\bibliographystyle{apsrev}

\bibliography{refgraph}

\begin{thebibliography}{44}
\expandafter\ifx\csname natexlab\endcsname\relax\def\natexlab#1{#1}\fi
\expandafter\ifx\csname bibnamefont\endcsname\relax
  \def\bibnamefont#1{#1}\fi
\expandafter\ifx\csname bibfnamefont\endcsname\relax
  \def\bibfnamefont#1{#1}\fi
\expandafter\ifx\csname citenamefont\endcsname\relax
  \def\citenamefont#1{#1}\fi
\expandafter\ifx\csname url\endcsname\relax
  \def\url#1{\texttt{#1}}\fi
\expandafter\ifx\csname urlprefix\endcsname\relax\def\urlprefix{URL }\fi
\providecommand{\bibinfo}[2]{#2}
\providecommand{\eprint}[2][]{\url{#2}}

\bibitem[{\citenamefont{Bukov et~al.}(2015)\citenamefont{Bukov, D'Alessio, and
  Polkovnikov}}]{BukovAdvPhys2015}
\bibinfo{author}{\bibfnamefont{M.}~\bibnamefont{Bukov}},
  \bibinfo{author}{\bibfnamefont{L.}~\bibnamefont{D'Alessio}},
  \bibnamefont{and}
  \bibinfo{author}{\bibfnamefont{A.}~\bibnamefont{Polkovnikov}},
  \bibinfo{journal}{Advances in Physics} \textbf{\bibinfo{volume}{64}},
  \bibinfo{pages}{139} (\bibinfo{year}{2015}).

\bibitem[{\citenamefont{Anderson et~al.}(2013)\citenamefont{Anderson, Spielman,
  and Juzeli\ifmmode~\bar{u}\else \={u}\fi{}nas}}]{AndersonPRL2013}
\bibinfo{author}{\bibfnamefont{B.~M.} \bibnamefont{Anderson}},
  \bibinfo{author}{\bibfnamefont{I.~B.} \bibnamefont{Spielman}},
  \bibnamefont{and}
  \bibinfo{author}{\bibfnamefont{G.}~\bibnamefont{Juzeli\ifmmode~\bar{u}\else
  \={u}\fi{}nas}}, \bibinfo{journal}{Phys. Rev. Lett.}
  \textbf{\bibinfo{volume}{111}}, \bibinfo{pages}{125301}
  (\bibinfo{year}{2013}).

\bibitem[{\citenamefont{Miyake et~al.}(2013)\citenamefont{Miyake, Siviloglou,
  Kennedy, Burton, and Ketterle}}]{MiyakePRL2013}
\bibinfo{author}{\bibfnamefont{H.}~\bibnamefont{Miyake}},
  \bibinfo{author}{\bibfnamefont{G.~A.} \bibnamefont{Siviloglou}},
  \bibinfo{author}{\bibfnamefont{C.~J.} \bibnamefont{Kennedy}},
  \bibinfo{author}{\bibfnamefont{W.~C.} \bibnamefont{Burton}},
  \bibnamefont{and} \bibinfo{author}{\bibfnamefont{W.}~\bibnamefont{Ketterle}},
  \bibinfo{journal}{Phys. Rev. Lett.} \textbf{\bibinfo{volume}{111}},
  \bibinfo{pages}{185302} (\bibinfo{year}{2013}).

\bibitem[{\citenamefont{Aidelsburger et~al.}(2013)\citenamefont{Aidelsburger,
  Atala, Lohse, Barreiro, Paredes, and Bloch}}]{AidelsburgerPRL2013}
\bibinfo{author}{\bibfnamefont{M.}~\bibnamefont{Aidelsburger}},
  \bibinfo{author}{\bibfnamefont{M.}~\bibnamefont{Atala}},
  \bibinfo{author}{\bibfnamefont{M.}~\bibnamefont{Lohse}},
  \bibinfo{author}{\bibfnamefont{J.~T.} \bibnamefont{Barreiro}},
  \bibinfo{author}{\bibfnamefont{B.}~\bibnamefont{Paredes}}, \bibnamefont{and}
  \bibinfo{author}{\bibfnamefont{I.}~\bibnamefont{Bloch}},
  \bibinfo{journal}{Phys. Rev. Lett.} \textbf{\bibinfo{volume}{111}},
  \bibinfo{pages}{185301} (\bibinfo{year}{2013}).

\bibitem[{\citenamefont{Zenesini et~al.}(2009)\citenamefont{Zenesini, Lignier,
  Ciampini, Morsch, and Arimondo}}]{ZenesiniPRL2009}
\bibinfo{author}{\bibfnamefont{A.}~\bibnamefont{Zenesini}},
  \bibinfo{author}{\bibfnamefont{H.}~\bibnamefont{Lignier}},
  \bibinfo{author}{\bibfnamefont{D.}~\bibnamefont{Ciampini}},
  \bibinfo{author}{\bibfnamefont{O.}~\bibnamefont{Morsch}}, \bibnamefont{and}
  \bibinfo{author}{\bibfnamefont{E.}~\bibnamefont{Arimondo}},
  \bibinfo{journal}{Phys. Rev. Lett.} \textbf{\bibinfo{volume}{102}},
  \bibinfo{pages}{100403} (\bibinfo{year}{2009}).

\bibitem[{\citenamefont{{Jotzu} et~al.}(2014)\citenamefont{{Jotzu}, {Messer},
  {Desbuquois}, {Lebrat}, {Uehlinger}, {Greif}, and
  {Esslinger}}}]{JotzuNAT2014}
\bibinfo{author}{\bibfnamefont{G.}~\bibnamefont{{Jotzu}}},
  \bibinfo{author}{\bibfnamefont{M.}~\bibnamefont{{Messer}}},
  \bibinfo{author}{\bibfnamefont{R.}~\bibnamefont{{Desbuquois}}},
  \bibinfo{author}{\bibfnamefont{M.}~\bibnamefont{{Lebrat}}},
  \bibinfo{author}{\bibfnamefont{T.}~\bibnamefont{{Uehlinger}}},
  \bibinfo{author}{\bibfnamefont{D.}~\bibnamefont{{Greif}}}, \bibnamefont{and}
  \bibinfo{author}{\bibfnamefont{T.}~\bibnamefont{{Esslinger}}},
  \bibinfo{journal}{Nature} \textbf{\bibinfo{volume}{515}},
  \bibinfo{pages}{237} (\bibinfo{year}{2014}).

\bibitem[{\citenamefont{Rechtsman et~al.}(2013)\citenamefont{Rechtsman, Zeuner,
  Plotnik, Lumer, Podolsky, Dreisow, Nolte, Segev, and Szameit}}]{rechtsman}
\bibinfo{author}{\bibfnamefont{M.~C.} \bibnamefont{Rechtsman}},
  \bibinfo{author}{\bibfnamefont{J.~M.} \bibnamefont{Zeuner}},
  \bibinfo{author}{\bibfnamefont{Y.}~\bibnamefont{Plotnik}},
  \bibinfo{author}{\bibfnamefont{Y.}~\bibnamefont{Lumer}},
  \bibinfo{author}{\bibfnamefont{D.}~\bibnamefont{Podolsky}},
  \bibinfo{author}{\bibfnamefont{F.}~\bibnamefont{Dreisow}},
  \bibinfo{author}{\bibfnamefont{S.}~\bibnamefont{Nolte}},
  \bibinfo{author}{\bibfnamefont{M.}~\bibnamefont{Segev}}, \bibnamefont{and}
  \bibinfo{author}{\bibfnamefont{A.}~\bibnamefont{Szameit}},
  \bibinfo{journal}{Nature} \textbf{\bibinfo{volume}{496}},
  \bibinfo{pages}{196} (\bibinfo{year}{2013}).

\bibitem[{\citenamefont{Goldman et~al.}(2016)\citenamefont{Goldman, Budich, and
  Zoller}}]{GoldmanNatPhys2016}
\bibinfo{author}{\bibfnamefont{N.}~\bibnamefont{Goldman}},
  \bibinfo{author}{\bibfnamefont{J.}~\bibnamefont{Budich}}, \bibnamefont{and}
  \bibinfo{author}{\bibfnamefont{P.}~\bibnamefont{Zoller}},
  \bibinfo{journal}{Nature Physics} \textbf{\bibinfo{volume}{12}},
  \bibinfo{pages}{639} (\bibinfo{year}{2016}).

\bibitem[{\citenamefont{Wilcox}(1967)}]{WilcoxJMP1967}
\bibinfo{author}{\bibfnamefont{R.}~\bibnamefont{Wilcox}},
  \bibinfo{journal}{Journal of Mathematical Physics}
  \textbf{\bibinfo{volume}{8}}, \bibinfo{pages}{962} (\bibinfo{year}{1967}).

\bibitem[{\citenamefont{Van-Brunt and Visser}(2015)}]{VanBruntJPA2015}
\bibinfo{author}{\bibfnamefont{A.}~\bibnamefont{Van-Brunt}} \bibnamefont{and}
  \bibinfo{author}{\bibfnamefont{M.}~\bibnamefont{Visser}},
  \bibinfo{journal}{Journal of Physics A: Mathematical and Theoretical}
  \textbf{\bibinfo{volume}{48}}, \bibinfo{pages}{225207}
  (\bibinfo{year}{2015}).

\bibitem[{\citenamefont{{Gritsev} and {Polkovnikov}}(2017)}]{GritsevArxiv2017}
\bibinfo{author}{\bibfnamefont{V.}~\bibnamefont{{Gritsev}}} \bibnamefont{and}
  \bibinfo{author}{\bibfnamefont{A.}~\bibnamefont{{Polkovnikov}}},
  \bibinfo{journal}{ArXiv e-prints}  (\bibinfo{year}{2017}),
  \eprint{1701.05276}.

\bibitem[{\citenamefont{Magnus}(1954)}]{Magnus1954}
\bibinfo{author}{\bibfnamefont{W.}~\bibnamefont{Magnus}},
  \bibinfo{journal}{Communications on Pure and Applied Mathematics}
  \textbf{\bibinfo{volume}{7}}, \bibinfo{pages}{649} (\bibinfo{year}{1954}),
  ISSN \bibinfo{issn}{1097-0312}.

\bibitem[{\citenamefont{Rahav et~al.}(2003)\citenamefont{Rahav, Gilary, and
  Fishman}}]{RahavPRA2003}
\bibinfo{author}{\bibfnamefont{S.}~\bibnamefont{Rahav}},
  \bibinfo{author}{\bibfnamefont{I.}~\bibnamefont{Gilary}}, \bibnamefont{and}
  \bibinfo{author}{\bibfnamefont{S.}~\bibnamefont{Fishman}},
  \bibinfo{journal}{Phys. Rev. A} \textbf{\bibinfo{volume}{68}},
  \bibinfo{pages}{013820} (\bibinfo{year}{2003}).

\bibitem[{\citenamefont{Mikami et~al.}(2016)\citenamefont{Mikami, Kitamura,
  Yasuda, Tsuji, Oka, and Aoki}}]{MikamiPRB2016}
\bibinfo{author}{\bibfnamefont{T.}~\bibnamefont{Mikami}},
  \bibinfo{author}{\bibfnamefont{S.}~\bibnamefont{Kitamura}},
  \bibinfo{author}{\bibfnamefont{K.}~\bibnamefont{Yasuda}},
  \bibinfo{author}{\bibfnamefont{N.}~\bibnamefont{Tsuji}},
  \bibinfo{author}{\bibfnamefont{T.}~\bibnamefont{Oka}}, \bibnamefont{and}
  \bibinfo{author}{\bibfnamefont{H.}~\bibnamefont{Aoki}},
  \bibinfo{journal}{Phys. Rev. B} \textbf{\bibinfo{volume}{93}},
  \bibinfo{pages}{144307} (\bibinfo{year}{2016}).

\bibitem[{\citenamefont{Ponte et~al.}(2015)\citenamefont{Ponte, Chandran,
  Papić, and Abanin}}]{PonteAOP2015}
\bibinfo{author}{\bibfnamefont{P.}~\bibnamefont{Ponte}},
  \bibinfo{author}{\bibfnamefont{A.}~\bibnamefont{Chandran}},
  \bibinfo{author}{\bibfnamefont{Z.}~\bibnamefont{Papić}}, \bibnamefont{and}
  \bibinfo{author}{\bibfnamefont{D.~A.} \bibnamefont{Abanin}},
  \bibinfo{journal}{Annals of Physics} \textbf{\bibinfo{volume}{353}},
  \bibinfo{pages}{196 } (\bibinfo{year}{2015}), ISSN \bibinfo{issn}{0003-4916}.

\bibitem[{\citenamefont{D'Alessio and Rigol}(2014)}]{LucaPRX2014}
\bibinfo{author}{\bibfnamefont{L.}~\bibnamefont{D'Alessio}} \bibnamefont{and}
  \bibinfo{author}{\bibfnamefont{M.}~\bibnamefont{Rigol}},
  \bibinfo{journal}{Phys. Rev. X} \textbf{\bibinfo{volume}{4}},
  \bibinfo{pages}{041048} (\bibinfo{year}{2014}).

\bibitem[{\citenamefont{Lazarides et~al.}(2014)\citenamefont{Lazarides, Das,
  and Moessner}}]{LazaridesPRE2014}
\bibinfo{author}{\bibfnamefont{A.}~\bibnamefont{Lazarides}},
  \bibinfo{author}{\bibfnamefont{A.}~\bibnamefont{Das}}, \bibnamefont{and}
  \bibinfo{author}{\bibfnamefont{R.}~\bibnamefont{Moessner}},
  \bibinfo{journal}{Phys. Rev. E} \textbf{\bibinfo{volume}{90}},
  \bibinfo{pages}{012110} (\bibinfo{year}{2014}).

\bibitem[{\citenamefont{Abanin et~al.}(2015)\citenamefont{Abanin, De~Roeck, and
  Huveneers}}]{AbaninPRL2015}
\bibinfo{author}{\bibfnamefont{D.~A.} \bibnamefont{Abanin}},
  \bibinfo{author}{\bibfnamefont{W.}~\bibnamefont{De~Roeck}}, \bibnamefont{and}
  \bibinfo{author}{\bibfnamefont{F.~m.~c.} \bibnamefont{Huveneers}},
  \bibinfo{journal}{Phys. Rev. Lett.} \textbf{\bibinfo{volume}{115}},
  \bibinfo{pages}{256803} (\bibinfo{year}{2015}).

\bibitem[{\citenamefont{Mori et~al.}(2016)\citenamefont{Mori, Kuwahara, and
  Saito}}]{MoriPRL2016}
\bibinfo{author}{\bibfnamefont{T.}~\bibnamefont{Mori}},
  \bibinfo{author}{\bibfnamefont{T.}~\bibnamefont{Kuwahara}}, \bibnamefont{and}
  \bibinfo{author}{\bibfnamefont{K.}~\bibnamefont{Saito}},
  \bibinfo{journal}{Phys. Rev. Lett.} \textbf{\bibinfo{volume}{116}},
  \bibinfo{pages}{120401} (\bibinfo{year}{2016}).

\bibitem[{\citenamefont{Kuwahara et~al.}(2016)\citenamefont{Kuwahara, Mori, and
  Saito}}]{KuwaharaAoP2016}
\bibinfo{author}{\bibfnamefont{T.}~\bibnamefont{Kuwahara}},
  \bibinfo{author}{\bibfnamefont{T.}~\bibnamefont{Mori}}, \bibnamefont{and}
  \bibinfo{author}{\bibfnamefont{K.}~\bibnamefont{Saito}},
  \bibinfo{journal}{Annals of Physics} \textbf{\bibinfo{volume}{367}},
  \bibinfo{pages}{96 } (\bibinfo{year}{2016}), ISSN \bibinfo{issn}{0003-4916}.

\bibitem[{\citenamefont{Prosen}(1998{\natexlab{a}})}]{prosen_98a}
\bibinfo{author}{\bibfnamefont{T.}~\bibnamefont{Prosen}},
  \bibinfo{journal}{Phys. Rev. Lett.} \textbf{\bibinfo{volume}{80}},
  \bibinfo{pages}{1808} (\bibinfo{year}{1998}{\natexlab{a}}).

\bibitem[{\citenamefont{Prosen}(2002)}]{ProsenPRE02}
\bibinfo{author}{\bibfnamefont{T.}~\bibnamefont{Prosen}},
  \bibinfo{journal}{Phys. Rev. E} \textbf{\bibinfo{volume}{65}},
  \bibinfo{pages}{036208} (\bibinfo{year}{2002}).

\bibitem[{\citenamefont{D’Alessio and Polkovnikov}(2013)}]{LucaAoP2013}
\bibinfo{author}{\bibfnamefont{L.}~\bibnamefont{D’Alessio}} \bibnamefont{and}
  \bibinfo{author}{\bibfnamefont{A.}~\bibnamefont{Polkovnikov}},
  \bibinfo{journal}{Annals of Physics} \textbf{\bibinfo{volume}{333}},
  \bibinfo{pages}{19 } (\bibinfo{year}{2013}), ISSN \bibinfo{issn}{0003-4916}.

\bibitem[{\citenamefont{Citro et~al.}(2015)\citenamefont{Citro, Dalla~Torre,
  D'Alessio, Polkovnikov, Babadi, Oka, and Demler}}]{citro_15}
\bibinfo{author}{\bibfnamefont{R.}~\bibnamefont{Citro}},
  \bibinfo{author}{\bibfnamefont{E.~G.} \bibnamefont{Dalla~Torre}},
  \bibinfo{author}{\bibfnamefont{L.}~\bibnamefont{D'Alessio}},
  \bibinfo{author}{\bibfnamefont{A.}~\bibnamefont{Polkovnikov}},
  \bibinfo{author}{\bibfnamefont{M.}~\bibnamefont{Babadi}},
  \bibinfo{author}{\bibfnamefont{T.}~\bibnamefont{Oka}}, \bibnamefont{and}
  \bibinfo{author}{\bibfnamefont{E.}~\bibnamefont{Demler}},
  \bibinfo{journal}{Annals of Physics} \textbf{\bibinfo{volume}{360}},
  \bibinfo{pages}{694} (\bibinfo{year}{2015}).

\bibitem[{\citenamefont{Lenarcic et~al.}(2017)\citenamefont{Lenarcic, Lange,
  and Rosch}}]{lenarcic_17}
\bibinfo{author}{\bibfnamefont{Z.}~\bibnamefont{Lenarcic}},
  \bibinfo{author}{\bibfnamefont{F.}~\bibnamefont{Lange}}, \bibnamefont{and}
  \bibinfo{author}{\bibfnamefont{A.}~\bibnamefont{Rosch}},
  \bibinfo{journal}{arXiv:1706.05700}  (\bibinfo{year}{2017}).

\bibitem[{\citenamefont{Weinberg and Bukov}(2017)}]{WeinbergQuSpin}
\bibinfo{author}{\bibfnamefont{P.}~\bibnamefont{Weinberg}} \bibnamefont{and}
  \bibinfo{author}{\bibfnamefont{M.}~\bibnamefont{Bukov}},
  \bibinfo{journal}{SciPost Phys.} \textbf{\bibinfo{volume}{2}},
  \bibinfo{pages}{003} (\bibinfo{year}{2017}).

\bibitem[{\citenamefont{Goldman and Dalibard}(2014)}]{GoldmanPRX2014}
\bibinfo{author}{\bibfnamefont{N.}~\bibnamefont{Goldman}} \bibnamefont{and}
  \bibinfo{author}{\bibfnamefont{J.}~\bibnamefont{Dalibard}},
  \bibinfo{journal}{Phys. Rev. X} \textbf{\bibinfo{volume}{4}},
  \bibinfo{pages}{031027} (\bibinfo{year}{2014}).

\bibitem[{\citenamefont{Cirac and Zoller}(1995)}]{CiracPRL1995}
\bibinfo{author}{\bibfnamefont{J.~I.} \bibnamefont{Cirac}} \bibnamefont{and}
  \bibinfo{author}{\bibfnamefont{P.}~\bibnamefont{Zoller}},
  \bibinfo{journal}{Phys. Rev. Lett.} \textbf{\bibinfo{volume}{74}},
  \bibinfo{pages}{4091} (\bibinfo{year}{1995}).

\bibitem[{\citenamefont{Lanyon et~al.}(2011)\citenamefont{Lanyon, Hempel, abd
  M.~M\"uller, Gerritsma, F.Zahringer, P.Schindler, J.T.Barreiro, Rambach,
  Kirchmair, Hennrich et~al.}}]{lanyon_11}
\bibinfo{author}{\bibfnamefont{B.~P.} \bibnamefont{Lanyon}},
  \bibinfo{author}{\bibfnamefont{C.}~\bibnamefont{Hempel}},
  \bibinfo{author}{\bibfnamefont{D.~N.} \bibnamefont{abd M.~M\"uller}},
  \bibinfo{author}{\bibfnamefont{R.}~\bibnamefont{Gerritsma}},
  \bibinfo{author}{\bibnamefont{F.Zahringer}},
  \bibinfo{author}{\bibnamefont{P.Schindler}},
  \bibinfo{author}{\bibnamefont{J.T.Barreiro}},
  \bibinfo{author}{\bibfnamefont{M.}~\bibnamefont{Rambach}},
  \bibinfo{author}{\bibfnamefont{G.}~\bibnamefont{Kirchmair}},
  \bibinfo{author}{\bibfnamefont{M.}~\bibnamefont{Hennrich}},
  \bibnamefont{et~al.}, \bibinfo{journal}{Science}
  \textbf{\bibinfo{volume}{334}}, \bibinfo{pages}{57} (\bibinfo{year}{2011}).

\bibitem[{\citenamefont{Blatt and Roos}(2012)}]{BlattNatPhys2012}
\bibinfo{author}{\bibfnamefont{R.}~\bibnamefont{Blatt}} \bibnamefont{and}
  \bibinfo{author}{\bibfnamefont{C.~F.} \bibnamefont{Roos}},
  \bibinfo{journal}{Nature Physics} \textbf{\bibinfo{volume}{8}},
  \bibinfo{pages}{277} (\bibinfo{year}{2012}).

\bibitem[{\citenamefont{Neill et~al.}(2016)\citenamefont{Neill, Roushan, Fang,
  Chen, Kolodrubetz, Chen, Megrant, Barends, Campbell, Chiaro
  et~al.}}]{neil_16}
\bibinfo{author}{\bibfnamefont{C.}~\bibnamefont{Neill}},
  \bibinfo{author}{\bibfnamefont{P.}~\bibnamefont{Roushan}},
  \bibinfo{author}{\bibfnamefont{M.}~\bibnamefont{Fang}},
  \bibinfo{author}{\bibfnamefont{Y.}~\bibnamefont{Chen}},
  \bibinfo{author}{\bibfnamefont{M.}~\bibnamefont{Kolodrubetz}},
  \bibinfo{author}{\bibfnamefont{Z.}~\bibnamefont{Chen}},
  \bibinfo{author}{\bibfnamefont{A.}~\bibnamefont{Megrant}},
  \bibinfo{author}{\bibfnamefont{R.}~\bibnamefont{Barends}},
  \bibinfo{author}{\bibfnamefont{B.}~\bibnamefont{Campbell}},
  \bibinfo{author}{\bibfnamefont{B.}~\bibnamefont{Chiaro}},
  \bibnamefont{et~al.}, \bibinfo{journal}{Nature Physics}
  \textbf{\bibinfo{volume}{12}}, \bibinfo{pages}{1037} (\bibinfo{year}{2016}).

\bibitem[{\citenamefont{Hoang et~al.}(2013)\citenamefont{Hoang, Gerving, Land,
  Anquez, Hamley, and Chapman}}]{hoang_13}
\bibinfo{author}{\bibfnamefont{T.~M.} \bibnamefont{Hoang}},
  \bibinfo{author}{\bibfnamefont{C.~S.} \bibnamefont{Gerving}},
  \bibinfo{author}{\bibfnamefont{B.~J.} \bibnamefont{Land}},
  \bibinfo{author}{\bibfnamefont{M.}~\bibnamefont{Anquez}},
  \bibinfo{author}{\bibfnamefont{C.~D.} \bibnamefont{Hamley}},
  \bibnamefont{and} \bibinfo{author}{\bibfnamefont{M.~S.}
  \bibnamefont{Chapman}}, \bibinfo{journal}{Phys. Rev. Lett.}
  \textbf{\bibinfo{volume}{111}}, \bibinfo{pages}{090403}
  (\bibinfo{year}{2013}).

\bibitem[{\citenamefont{Gorbatsevich et~al.}(1997)\citenamefont{Gorbatsevich,
  Onishchik, and Vinberg}}]{vv1997foundations}
\bibinfo{author}{\bibfnamefont{V.}~\bibnamefont{Gorbatsevich}},
  \bibinfo{author}{\bibfnamefont{A.~L.} \bibnamefont{Onishchik}},
  \bibnamefont{and} \bibinfo{author}{\bibfnamefont{E.~B.}
  \bibnamefont{Vinberg}}, \emph{\bibinfo{title}{Foundations of Lie theory and
  Lie transformation groups}} (\bibinfo{publisher}{Springer},
  \bibinfo{year}{1997}).

\bibitem[{\citenamefont{Untidt and Nielsen}(2002)}]{UntidtPRE2002}
\bibinfo{author}{\bibfnamefont{T.~S.} \bibnamefont{Untidt}} \bibnamefont{and}
  \bibinfo{author}{\bibfnamefont{N.~C.} \bibnamefont{Nielsen}},
  \bibinfo{journal}{Phys. Rev. E} \textbf{\bibinfo{volume}{65}},
  \bibinfo{pages}{021108} (\bibinfo{year}{2002}).

\bibitem[{\citenamefont{Scharf}(1988)}]{ScharfJPA1988}
\bibinfo{author}{\bibfnamefont{R.}~\bibnamefont{Scharf}},
  \bibinfo{journal}{Journal of Physics A: Mathematical and General}
  \textbf{\bibinfo{volume}{21}}, \bibinfo{pages}{2007} (\bibinfo{year}{1988}).

\bibitem[{\citenamefont{M{\'e}zard et~al.}(1987)\citenamefont{M{\'e}zard,
  Parisi, and Virasoro}}]{Mezard1987spin}
\bibinfo{author}{\bibfnamefont{M.}~\bibnamefont{M{\'e}zard}},
  \bibinfo{author}{\bibfnamefont{G.}~\bibnamefont{Parisi}}, \bibnamefont{and}
  \bibinfo{author}{\bibfnamefont{M.}~\bibnamefont{Virasoro}},
  \emph{\bibinfo{title}{Spin glass theory and beyond: An Introduction to the
  Replica Method and Its Applications}}, vol.~\bibinfo{volume}{9}
  (\bibinfo{publisher}{World Scientific Publishing Co Inc},
  \bibinfo{year}{1987}).

\bibitem[{\citenamefont{Engel and Van~den Broeck}(2001)}]{engel2001statistical}
\bibinfo{author}{\bibfnamefont{A.}~\bibnamefont{Engel}} \bibnamefont{and}
  \bibinfo{author}{\bibfnamefont{C.}~\bibnamefont{Van~den Broeck}},
  \emph{\bibinfo{title}{Statistical mechanics of learning}}
  (\bibinfo{publisher}{Cambridge University Press}, \bibinfo{year}{2001}).

\bibitem[{\citenamefont{Cardy et~al.}(2008)\citenamefont{Cardy,
  Castro-Alvaredo, and Doyon}}]{CardyJSP2008}
\bibinfo{author}{\bibfnamefont{J.~L.} \bibnamefont{Cardy}},
  \bibinfo{author}{\bibfnamefont{O.~A.} \bibnamefont{Castro-Alvaredo}},
  \bibnamefont{and} \bibinfo{author}{\bibfnamefont{B.}~\bibnamefont{Doyon}},
  \bibinfo{journal}{Journal of Statistical Physics}
  \textbf{\bibinfo{volume}{130}}, \bibinfo{pages}{129} (\bibinfo{year}{2008}),
  ISSN \bibinfo{issn}{1572-9613}.

\bibitem[{sup()}]{suppmat}
\bibinfo{note}{See Supplemental Material at [URL will be inserted by
  publisher].}

\bibitem[{\citenamefont{Prosen}(1998{\natexlab{b}})}]{ProsenJPA1998}
\bibinfo{author}{\bibfnamefont{T.}~\bibnamefont{Prosen}},
  \bibinfo{journal}{Journal of Physics A: Mathematical and General}
  \textbf{\bibinfo{volume}{31}}, \bibinfo{pages}{L397}
  (\bibinfo{year}{1998}{\natexlab{b}}).

\bibitem[{Note1()}]{Note1}
Note1, \bibinfo{note}{although $\Delta _n$ scales linearly with the system size
  $L$, this accounts for expectation values of nonlocal operators. The accuracy
  of the dynamics of local observables is independent of system size \cite
  {suppmat}.}

\bibitem[{Note2()}]{Note2}
Note2, \bibinfo{note}{we give a more rigorous discussion and estimate about the
  magnitudes of $\Gamma _r$ in a subsequent publication.}

\bibitem[{\citenamefont{Khinchin}(1997)}]{khinchin1997continued}
\bibinfo{author}{\bibfnamefont{A.~Y.} \bibnamefont{Khinchin}},
  \emph{\bibinfo{title}{Continued fractions}} (\bibinfo{publisher}{Dover
  publications}, \bibinfo{year}{1997}).

\bibitem[{\citenamefont{Arnold}(2012)}]{arnold2012geometrical}
\bibinfo{author}{\bibfnamefont{V.~I.} \bibnamefont{Arnold}},
  \emph{\bibinfo{title}{Geometrical methods in the theory of ordinary
  differential equations}}, vol. \bibinfo{volume}{250}
  (\bibinfo{publisher}{Springer Science \& Business Media},
  \bibinfo{year}{2012}).

\end{thebibliography}
 
\onecolumngrid 
\appendix
\section{Supplementary material for "Replica resummation of the Baker-Campbell-Hausdorff series"}
\setcounter{equation}{0}
\renewcommand{\theequation}{S\arabic{equation}}

\setcounter{figure}{0}
\renewcommand{\thefigure}{S\arabic{figure}}
\section{Evaluation of the replica limit in the first order correction}
The first order correction in the replica trick from Eq.~\eqref{eq:Gamma_noncomm} is simply written as
\begin{align}\label{eq:Gamma_1_gen}
 \Gamma_1=\mL\sum_{0\leq m <\rho}  \Hm{m}=\mL \sum_{0\leq m <\rho} (U_0^{\text{ad}})^m H_1=\lim_{\rho\rightarrow 0}\frac{1}{\rho} \frac{(U_0^{\text{ad}})^{\rho}-1}{U_0^{\text{ad}}-1}=\frac{\log(U_0^{\text{ad}})}{U_0^{\text{ad}}-1} 
\end{align}
where we expressed $\Hm{m}$ by the adjoint action of $U_0$ as $\Hm{m}=U_0^{-m}H_1 U_0^{m}=(U_0^{\text{ad}})^m H_1$ and the substitution of $U_0^{\text{ad}}=e^{i J\text{ad}_{H_0}}$ brings us to Eq.~\eqref{eq:BCH_resum_ord1} of the main text. 
\section{Nested commutator expression of replica expansion}
A direct evaluation of the series expansion of the integer powers of the time evolution operator leads to Eq.~\eqref{eq:Gamma_noncomm} of the main text. However, similar to the BCH expansion, it is expected that the corrections can be written in terms of nested commutators, which produce local operators if the original Hamiltonians are local. We denote 
the sums in Eq.~\eqref{eq:Gamma_noncomm} by $\eta_r$ before evaluating the replica limit, that is, $\Gamma_r=\frac{(-i)^{r-1}}{r!}\mL \eta_r$, and
\begin{align}\label{eqA:eta}
 \eta_r=\sum_{0\leq m_1\leq \dots m_r<\rho} \hspace{-1em} \Hm{m_r}\Hm{m_{r-1}}\dots \Hm{m_1} c_{m_1 \dots m_r} \,.
\end{align}
The corresponding nested commutator expressions from Eq.~\eqref{eq:Gamma_comm} are denoted by $\tilde{\eta}_r$,
\begin{align}
\tilde{\eta}_r=\sum_{0\leq m_1\leq \dots m_r<\rho} \hspace{-1em} [\Hm{m_r},\dots [\Hm{m_2},\Hm{m_1}]]] c_{m_2 \dots m_r}\,.
\end{align}
We show that $\eta_r$ can be expressed as a sum of $\tilde{\eta_r}$ and terms which vanish in the replica limit, that is, $\mL \eta_r=\mL \tilde{\eta}_r$. 
A necessary condition for the replica trick to work up to the $r^{\text{th}}$ order is to have an analytical continuation for which $\eta_r=\mO(\rho)$. This means that polynomials of at least second order of $\{\eta_1,\dots,\eta_{r-1}\}$ vanish in the replica limit. The second order correction was discussed in the main text, in this notation 
\begin{align}
 \tilde{\eta}_2&=\sum_{0 \leq m_1 \leq  m_2<\rho} [\Hm{m_2},\Hm{m_1}]\\
 \eta_1^2 &=\sum_{0 \leq m_1 \leq  m_2<\rho} \Hm{m_2}\Hm{m_1}+\sum_{0 \leq m_1 <  m_2<\rho} \Hm{m_1}\Hm{m_2}\\
 \eta_2 &=\sum_{0 \leq m_1 \leq  m_2<\rho} \Hm{m_2}\Hm{m_1}(2-\delta_{m_1,m_2})= \tilde{\eta}_2 +\eta_1^2
\end{align}
Below we list the combinations of the operators $\eta_s$ which produce the nested commutators in Eq.~\eqref{eq:Gamma_comm} of the main text, up to the $5^{\text{th}}$ order (note the non-commutativity of the different $\eta_s$s). 
\begin{align}
 \tilde{\eta}_1=&\eta_1\\
 \tilde{\eta}_2=&\eta_2-\eta_1^2\\
 \tilde{\eta}_3=&\eta_3-2 \eta_2 \eta_1-\eta_1 \eta_2+2\eta_1^3\\
 \tilde{\eta}_4=&\eta_4-3\eta_3 \eta_1-\eta_1 \eta_3-3 \eta_2^2+6\eta_2 \eta_1^2+3 \eta_1 \eta_2 \eta_1+3 \eta_1^2\eta_2-6\eta_1^4\\
 \tilde{\eta}_5=&\eta_5-4\eta_4 \eta_1-\eta_1 \eta_4-6\eta_3 \eta_2-4\eta_2 \eta_3+12\eta_3 \eta_1^2 +4\eta_1 \eta_3 \eta_1+4\eta_1^2 \eta_3+12\eta_2^2 \eta_1+12\eta_2 \eta_1 \eta_2 \nonumber  \\ & +6 \eta_1 \eta_2^2 -24\eta_2 \eta_1^3-12\eta_1 \eta_2 \eta_1^2 -12 \eta_1^2 \eta_2 \eta_1-12\eta_1^3 \eta_2+24\eta_1^5
 \end{align}
We note that if the operators $\eta_s$ commute then these equations reduce to the cumulant expansion.

\section{Extraction of the BCH coefficients from the replica expansion}
It is easy to check that series expansion, or equivalently, the partial derivatives of the corrections $\Gamma_r$ at $J=0$ reproduce the commutators from the BCH formula containing $r$ instances of $H_1$, e.g. 
\begin{align}
\partial_J \Gamma_1\big|_{J=0}  &= \ct \sum_j \sigma_j^y\sigma_{j+1}^z+\sigma_{j}^z\sigma_{j+1}^y=-\frac{1}{2}i [H_0,H_1]\\
\frac{1}{2}\partial_J^2 \Gamma_1\big|_{J=0}  &= -\frac{2}{3}\ct \sum_j \sigma_j^x+\sigma_j^z\sigma_{j+1}^x\sigma_{j+2}^z=-\frac{1}{12}[H_0,[H_0,H_1]]\\
\partial_J \Gamma_2\big|_{J=0}  &= \frac{2}{3}\ct \sum_j \st (\sigma_j^x \sigma_{j+1}^z+\sigma_j^z \sigma_{j+1}^x)+   2\ct   (\sigma_j^y \sigma_{j+1}^y-\sigma_j^z \sigma_{j+1}^z)       = -\frac{1}{12}[H_1,[H_1,H_0]]  \\             
\frac{1}{2}\partial_J^2 \Gamma_2\big|_{J=0}  &= \frac{2}{3}\ct \sum_j \st (\sigma_j^y +\sigma_{j}^z\sigma_{j+1}^y \sigma_{j+2}^z) -  \ct   (\sigma_j^z \sigma_{j+1}^x \sigma_{j+2}^y+\sigma_j^y \sigma_{j+1}^x \sigma_{j+2}^z)       = -i\frac{1}{24}[H_1,[H_0,[H_0,H_1]]] 
\end{align}

\section{Expectation value of local operators}
\begin{figure}[htb!]
\centering
\includegraphics[width=10.0cm]{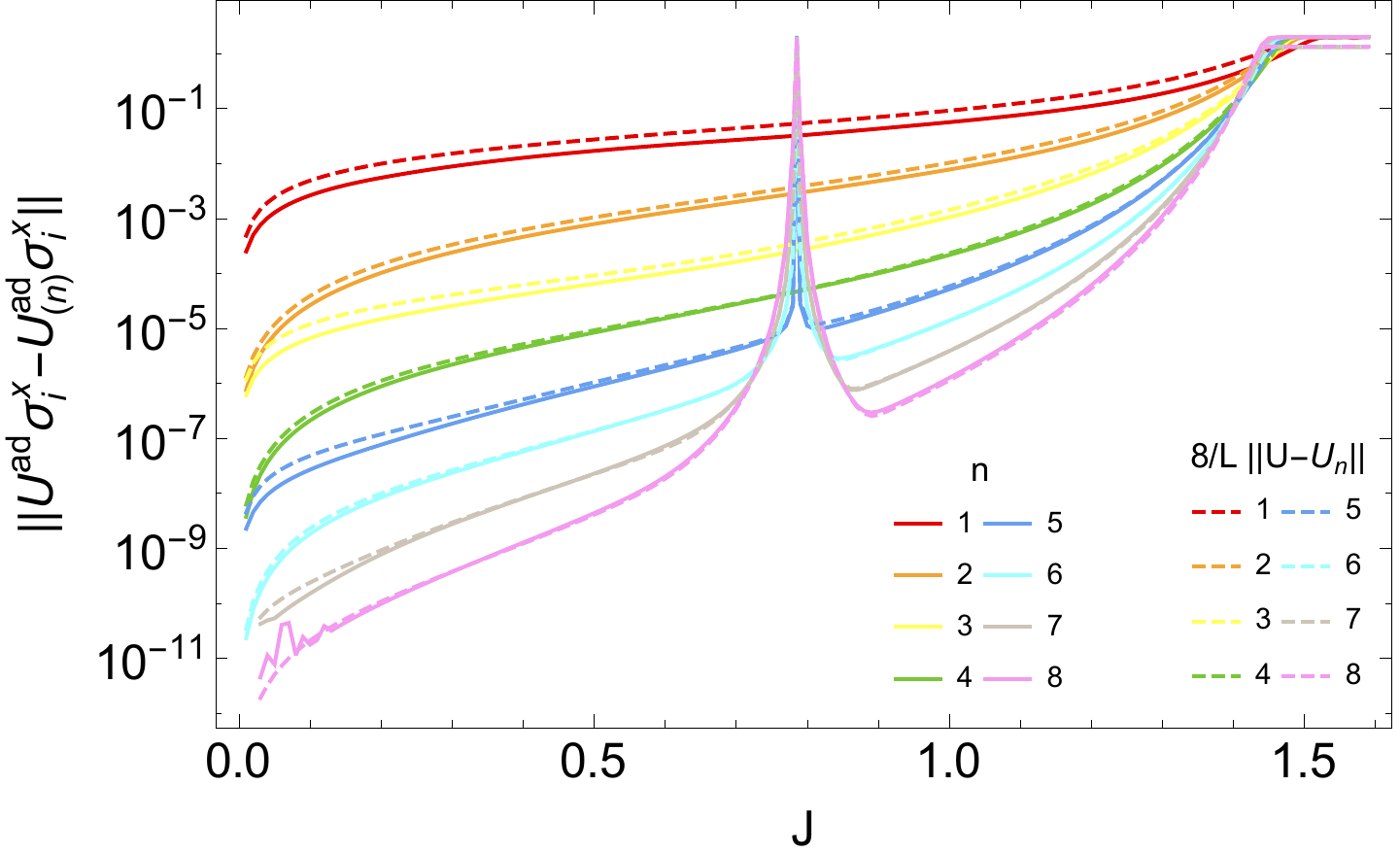}
\caption{Accuracy of the dynamics generated by the replica Floquet Hamiltonian in the kicked tilted field Ising model (solid lines), with parameters $h=0.1$, $\ct=0.8$, $\st=0.6$, $L=12$. The scale of the accuracy $\tilde{\Delta}(\sigma_i^{x})$ is well captured by an intensive ($1/L$ scaled) norm difference ${\Delta}_n/L$ (dashed curves). }
\label{fig:normdiff_local}
\end{figure}

An important measure of the accuracy of the replica expansion is how well it reproduces the dynamics of local observables. This can be quantified by the operator norm of the difference between the time evolved observable with respect to the exact and approximate time evolution: $\tilde{\Delta}_n(A)=\|U^{\text{ad}}A-U^{\text{ad}}_{(n)}A\|$, for a local operator $A$ and notation $U^{\text{ad}}A \equiv U^{-1}A U$. This measure is independent of the system size up to an exponentially small correction due to Lieb-Robinson bounds. In principle it could be possible that the resonances seen in Figure~\ref{fig:replica_vs_BCH} correspond to some non-local observables and they do not affect the dynamics of local operators. In contrast, we find that the accuracy measured by the time evolution of single spin operators $\sigma_i^{\mu}$, $\mu\in \{x,y,z\}$ follows closely the estimate given by $\Delta_n$, as shown in Figure~\ref{fig:normdiff_local} for $\sigma_i^{x}$. The other spin components behave in qualitatively the same way.  
As discussed in the main text, the spectral norm of the difference of the exact and approximate time evolution operators $\Delta_n=\|U-U^{(n)}\|$ bounds the accuracy of the expectation value of any dynamical observable $A$, including non-local ones, as $\tilde{\Delta}_n(A)\leq (2 \Delta_n +\Delta_n^2)\|A\|$. However, to account for the possibly non-local observables, $\Delta_n$ scales linearly with system size, and underestimates the accuracy of the replica expansion for the time evolution of local operators. 

\section{Hilbert-Schmidt norm of the corrections}
\begin{figure}[htb!]
\centering
\includegraphics[height=5.5cm]{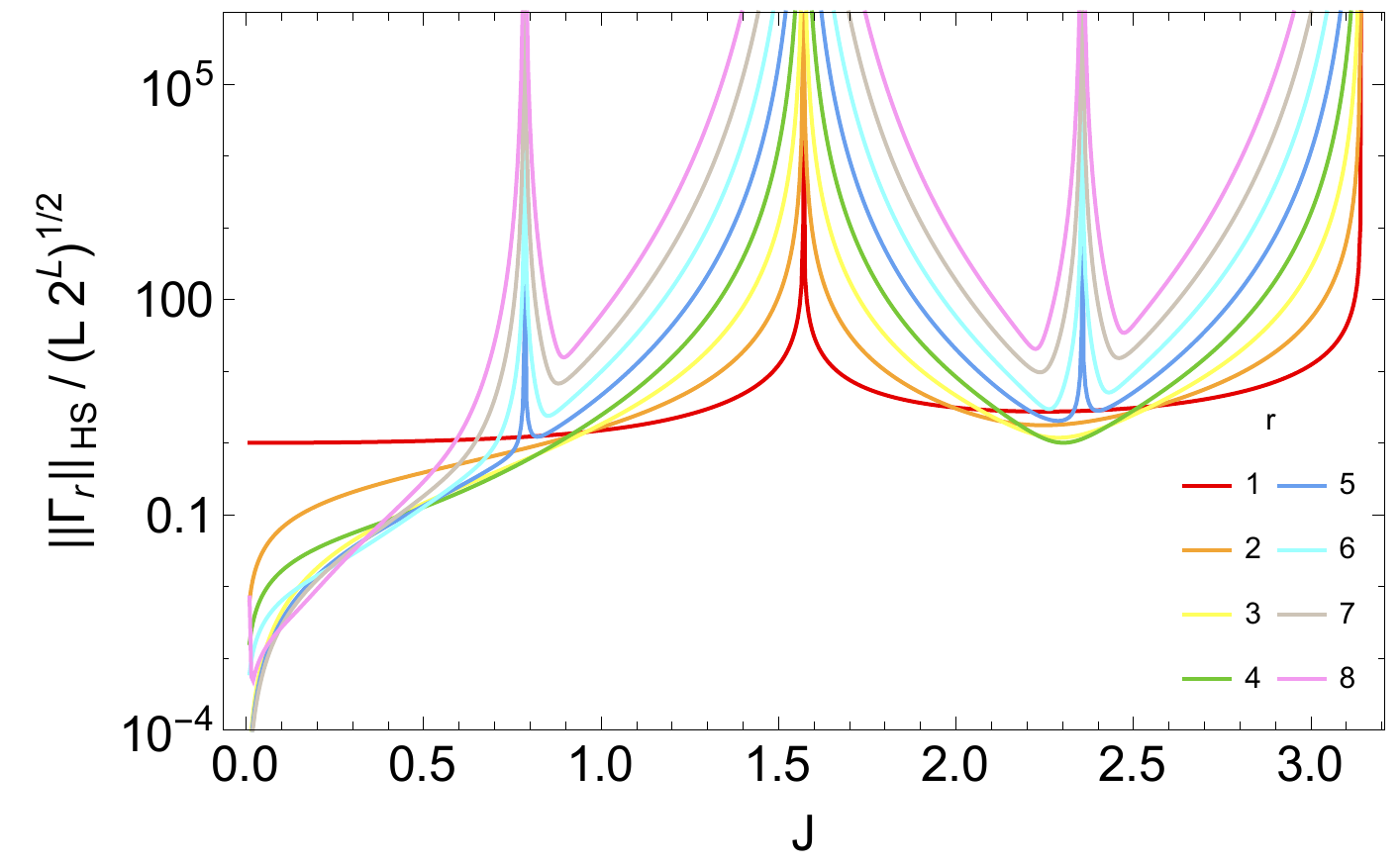}\includegraphics[height=5.5cm]{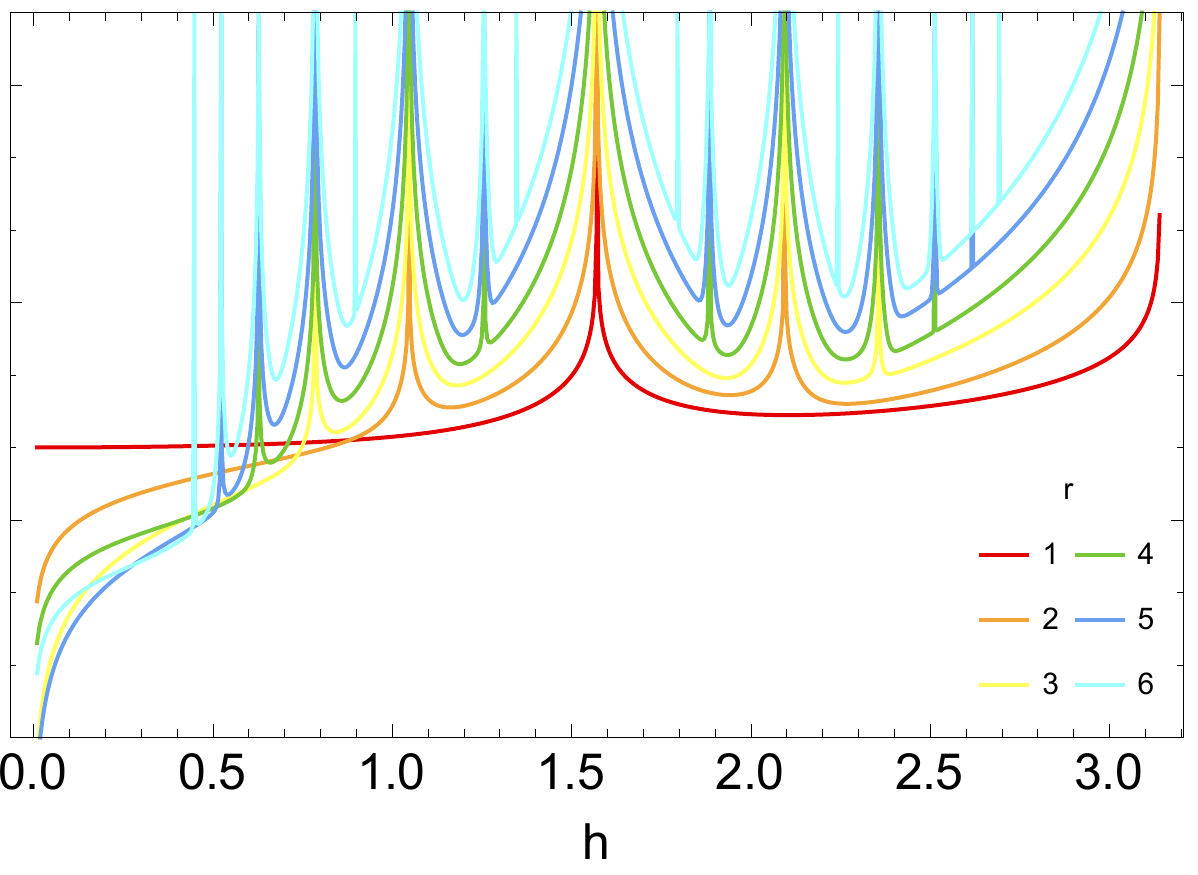}
\caption{Hilbert-Schmidt norm of the corrections $\Gamma_r$ in the kicked Ising model, the kick is the tilted magnetic field in (a) and the Ising interaction in (b). The parameters are $\ct=0.8$, $\st=0.6$.}
\label{fig:frobnorm}
\end{figure}

\begin{figure}[htb!]
\centering
\includegraphics[width=8.5cm]{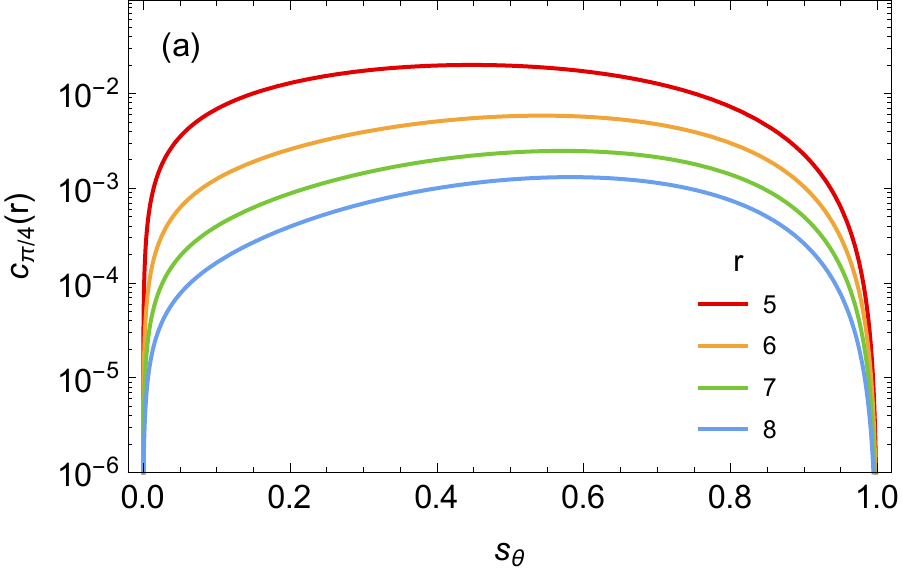}\includegraphics[width=8.5cm]{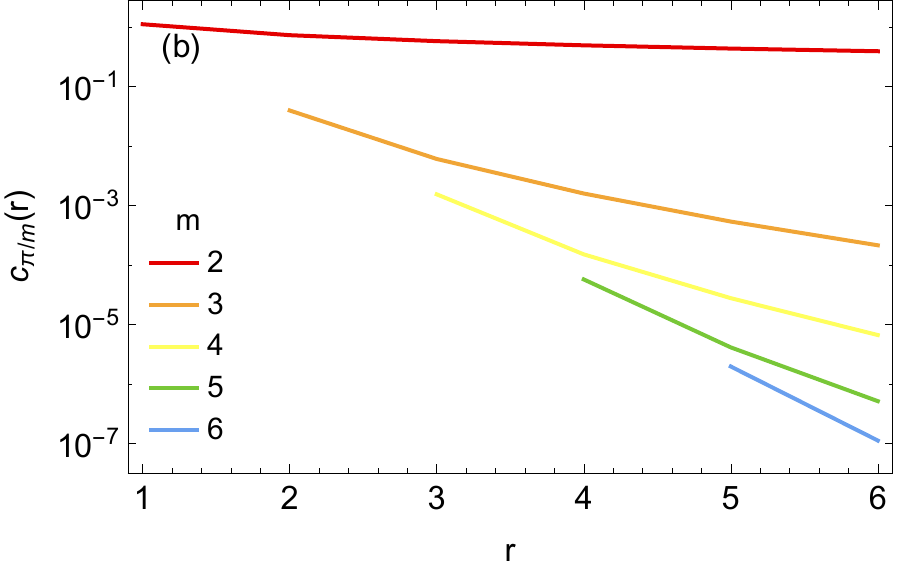}
\caption{(a) The coefficients of the divergence in Eq.~\eqref{eq:div_coeff} at $J=\pi/4$ as a function of the angle characterizing the kicking magnetic field. The resonance is absent in the first four orders as well as in the integrable $\st=0$ and $\st=1$ limits. (b) The same coefficients in the case of interaction kick at various resonances indexed by $m$, at parameters $\ct=0.8$, $\st=0.6$. Note that these numerical data correspond to interaction kicks, since in this regime we obtain a new resonance at each consecutive order, and that the expansion parameter is $J$ rather than $h$.}
\label{fig:frobnorm_coeff}
\end{figure}

The Hilbert-Schmidt norm of corrections $\Gamma_r$ are illustrated in Figure~\ref{fig:frobnorm} for both magnetic field and interaction kicks. Note that in the latter case the expansion parameter is $J$ rather than $h$.
The asymptotic behavior of the Hilbert-Schmidt norm near divergencies can be further analyzed, as in Eq.~\eqref{eq:div_coeff} of the main text,
\begin{align} \label{eq:div_coeff_supp}
f^r_{J^*}(\delta J)&=\frac{c_{J^*}(r)}{\delta J^{r-r_{J^*}}}+\mathcal{O}(\delta J^{-(r-r_{J^*})+1})\\
f^r_{h^*}(\delta h)&=\frac{c_{h^*}(r)}{\delta h^{r-r_{h^*}}}+\mathcal{O}(\delta h^{-(r-r_{h^*})+1})\,,
\end{align}
where $r_{J^*}+1$ ($r_{h^*}+1$) determines the order of expansion at which the divergence at coupling $J=J^*$ ($h=h^*$) first appears. The coefficient $c_{\pi/4}(r)$ is plotted in Figure~\ref{fig:frobnorm_coeff} (a) as a function of $\st$ for kicking magnetic field. It is more informative to analyze the model with interaction kicks, because we have much more resonances at hand in this situation. The coefficients $c_{h^*}(r)$ are shown for that model in Figure~\ref{fig:frobnorm_coeff} (b). The new resonances appearing at higher orders are characterized by smaller coefficients than the previous ones, at least up to the $6^{\text{th}}$ order.

\section{Periodicity preserving analytical continuation}
In the manuscript we used the naive analytical continuation, that is, we evaluated the replica sums in Eq.~\eqref{eq:replica_sum} with the prescription Eq.~\eqref{eq:repl_sum_part}, then we continued the resulting expression to arbitrary real values of the replica index $\rho$. This results in replica coefficients continuous in $J$ except for the resonances. The denominators in Eq.~\eqref{eq:repl_sum_part} do not break the $2\pi$ periodicity of the time evolution operator $U$ in variable $J$. The linear in $J$ factors originate in the replica limit of the expressions $e^{i 4 J z \rho}-1$,
\begin{align}
\mL e^{i 4 J z \rho}-1= \lim_{\rho \rightarrow 0}\frac{e^{i 4 J z \rho}-1}{\rho}=\log e^{i 4 J z \rho}=i m_z(4 J z)
\end{align}
where $z$ is an integer. In the naive analytical continuation, for all values of $z$, we chose $m_z$ as the identity function. However, we can choose a different branch cut of the logarithm, which is equivalent to dividing our analytical continued function by $e^{i 2\pi \zeta \rho}$, which is unity at integer values of $\rho$, and $\zeta$ is an integer defining the new analytical continuation. In principle we are allowed to define a different $\zeta$ for every single term in the expansion, here we consider it as a function of $z$. The periodicity $H_F(J)=H_F(J+2\pi)$ can be enforced by substituting $J$ by $J\, mod\, 2\pi$ in the replica coefficients. This corresponds to the analytical continuation defined by $m_z(4 J z)=4 z (J\, mod\, 2\pi)$.

From the point of view of observables, the periodicity in $J$ is shorter, since $U$ only obtains a real phase factor $(\pm 1)$ following a shift of $\pi$:  $U(J+\pi)=(-1)^{b} U(J)$, where $b$ is the number of nearest neighbor bonds in the system. In open boundary conditions $b=L-1$ and in periodic boundary conditions (PBC) $b=L$. The period is further decreased in PBC, because then $U(J+\pi/2)=i^L U(J)$ since
\begin{align}
e^{i \pi/2 \sigma_i^z\sigma_{i+1}^z}=i \sigma_i^z\sigma_{i+1}^z 
\end{align}
and $\prod \sigma_i^z\sigma_{i+1}^z$ is the identity operator if PBC is applied. Thus, at appropriate system sizes, the period of $U$ in $J$ is halved or even quartered, which can be similarly established by a corresponding analytical continuation, that is, by the proper choice of the function $m_z$.

\end{document}